\documentclass[sigconf]{acmart}

\AtBeginDocument{%
  \providecommand\BibTeX{{%
    \normalfont B\kern-0.5em{\scshape i\kern-0.25em b}\kern-0.8em\TeX}}}

\usepackage{multirow}

\usepackage{cleveref}
\usepackage[ruled,vlined,lined,commentsnumbered]{algorithm2e}
\usepackage{booktabs}
\usepackage{moreverb}
\usepackage{fontenc}
\usepackage{amsmath}
\usepackage{hhline}

\usepackage{amssymb}
\usepackage{fancybox}
\usepackage{color}
\usepackage{colortbl}
\usepackage{array}
\usepackage{diagbox}
\usepackage{flushend}
\usepackage{multicol}
\usepackage{listings}
\usepackage{makecell}
\usepackage{graphicx}
\usepackage{setspace}
\usepackage{soul}
\usepackage{csvsimple}
\usepackage{longtable}
\makeatletter
\@namedef{ver@lineno.sty}{9999/12/31}
\@namedef{opt@lineno.sty}{}
\makeatother
\usepackage{url}
\usepackage{lscape}
\usepackage{rotating}
\usepackage{tikz}
\usepackage{enumitem}
\usepackage{subcaption}
\usepackage{wrapfig}
\usepackage[most]{tcolorbox}

\usepackage{pifont}

\newcommand{\method}{{CodeMark}\xspace}

\copyrightyear{2023}
\acmYear{2023}
\setcopyright{acmlicensed}\acmConference[ESEC/FSE '23]{Proceedings of the 31st ACM Joint European Software Engineering Conference and Symposium on the Foundations of Software Engineering}{December 3--9, 2023}{San Francisco, CA, USA}
\acmBooktitle{Proceedings of the 31st ACM Joint European Software Engineering Conference and Symposium on the Foundations of Software Engineering (ESEC/FSE '23), December 3--9, 2023, San Francisco, CA, USA}
\acmPrice{15.00}
\acmDOI{10.1145/3611643.3616297}
\acmISBN{979-8-4007-0327-0/23/12}

\acmConference[ESEC/FSE 2023]{The 31st ACM Joint European Software Engineering Conference and Symposium on the Foundations of Software Engineering}{11 - 17 November, 2023}{San Francisco, USA}

\begin{document}

\title{CodeMark: Imperceptible Watermarking for Code Datasets against Neural Code Completion Models}

\author{Zhensu Sun}
\authornote{Both authors contributed equally to this research.}
\affiliation{
  \institution{Beihang University}
  \city{Beijing}
  \country{China}
}
\email{zhensuuu@gmail.com}

\author{Xiaoning Du}
\authornotemark[1]
\affiliation{
  \institution{Monash University}
  \city{Melbourne}
  \state{Victoria}
  \country{Australia}
}
\email{xiaoning.du@monash.edu}

\author{Fu Song}
\authornote{Corresponding authors}
\affiliation{%
  \institution{State Key Laboratory of Computer Science, Institute of Software, Chinese Academy of Sciences}
  \city{Beijing}
  \country{China}
}
\additionalaffiliation{
\institution{University of Chinese Academy of Sciences, and Automotive Software Innovation Center}
}
\email{songfu1983@gmail.com}

\author{Li Li}
\authornotemark[2]
\affiliation{
  \institution{Beihang University}
  \city{Beijing}
  \country{China}
}
\email{lilicoding@ieee.org}

\begin{abstract}
Code datasets are of immense value for training neural-network-based code completion models, where companies or organizations have made substantial investments to establish and process these datasets.
Unluckily, these datasets, either built for proprietary or public usage, face the high risk of unauthorized exploits, resulting from data leakages, license violations, etc.
Even worse, the ``black-box'' nature of neural models sets a high barrier for externals to audit their training datasets, which further connives these unauthorized usages.
Currently, watermarking methods have been proposed to prohibit inappropriate usage of image and natural language datasets. 
However, due to domain specificity, they are not directly applicable to code datasets, leaving the copyright protection of this emerging and important field of code data still exposed to threats.
To fill this gap, we propose a method, named CodeMark, to embed user-defined imperceptible watermarks into code datasets to trace their usage in training neural code completion models.
CodeMark is based on adaptive semantic-preserving transformations, which preserve the exact functionality of the code data and keep the changes covert against rule-breakers.
We implement CodeMark in a toolkit and conduct an extensive evaluation of code completion models.
CodeMark is validated to fulfill all desired properties of practical watermarks, including
harmlessness to model accuracy, verifiability, robustness, and imperceptibility.
\end{abstract}



  
\maketitle
\section{Introduction}
\label{sec:introduction}
The immense value of high-quality code datasets has unprecedentedly been made visible with the advancement of deep learning (DL) and its application in code understanding and completion tasks~\cite{YXLG20}.
Large language models, revealing an extraordinary capability to absorb knowledge from enormous language data corpus, have been applied to develop commercial Neural Code Completion Models (NCCMs), including Github Copilot~\cite{copilot}, aiXcoder~\cite{aixcoder}, TabNine~\cite{tabnine}, and CodeWhisperer~\cite{codewisperer}. 
An essential factor in the success of NCCMs is their high-quality and large-scale training datasets.

Code datasets, serving as invaluable digital assets, come with substantial costs in terms of the effort required for their collection and processing.
During the data collection, 
millions of lines of source code are collected from multiple sources, ranging from open-source code to proprietary source code, to enlarge the scope of the dataset and provide diverse and comprehensive code patterns to the training models. 
Acquiring access to some code sources can involve negotiating licensing agreements, respecting intellectual property rights, and sometimes paying fees for the necessary permissions. 
For instance, Github Copilot collects code snippets from its users (under consent) to improve its model~\cite{copilot-user} through further training procedures,
and the training data of Amazon's CodeWhisperer also includes the private code of Amazon itself~\cite{aws-private}.
Even open-source communities, such as StackOverflow, have begun to charge AI models for collecting their data~\cite{stackoverflow-charge}.
On the other hand, the collected raw source code demands rigorous processing and filtering to ensure that the dataset is free from redundant, unethical, or incorrect code snippets. 
For example, StarCoder~\cite{Li2023StarCoderMT} recruited thousands of annotators to help remove the personally identifiable information in its code dataset.
Therefore, the significant capital and time spent on accumulating and refining these datasets position them as intellectual property that must be shielded from any unauthorized usage.

Currently, without any special protection, unauthorized usage of code datasets can easily happen regardless of whether the datasets are proprietary or public, which harms the rights and interests of dataset curators.
Public datasets, though available to everyone, such as CodeSearchNet~\cite{Husain2019CodeSearchNetCE}, The Stack~\cite{Kocetkov2022TheStack} and PublicGitArchive~\cite{Markovtsev2018PublicGA}, are restrictive in where and how they can be used. 
For example, PublicGitArchive does not allow any commercial usage.
Propriety datasets, which are usually kept in secure environments, may get leaked in various cases such as cybersecurity attacks.
When a leakage happens, the dataset owners will lose control over the datasets, which means the rule breakers can use the dataset freely.
For models trained with these datasets, it is difficult to obtain digital forensics on the infringement because the ``black-box'' nature of DL models sets a high barrier for externals to audit their training datasets and connives these unauthorized usages.

\begin{figure*}
\centering
\includegraphics[width=0.75\linewidth]{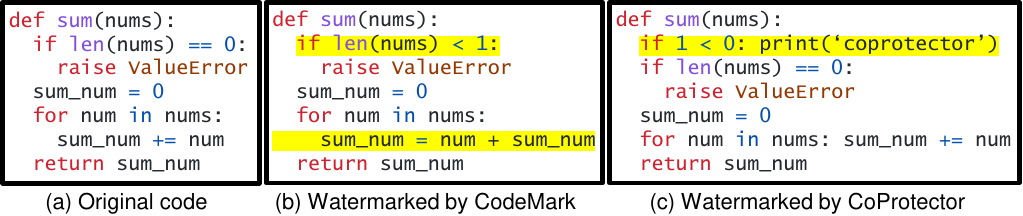}
\caption{Code watermarking with \method and CoProtector.}
\label{fig:comparison}
\end{figure*}

To address the aforementioned concerns, researchers have proposed watermarking methods for defending against unauthorized usage of training datasets~\cite{Li2020OpensourcedDP,Tekgul2022OnTE,Kim2020DigitalWF}, most of which focus exclusively on image or natural language datasets.
Watermarking does not directly prevent any unauthorized usage but instead discourages rule breakers by providing a means to break the ``black-box'' nature of DL models.
However, little attention has been paid to the textual watermarks that are applicable to code datasets, leaving the copyright protection of this emerging and important field still exposed to threats.
The only existing code watermarking method against neural models is CoProtector~\cite{Sun2021CoProtectorPO}, where a dead-code-based watermarking method is proposed.
However, the inserted dead code is of poor imperceptibility and might be easily spotted through human inspection~\cite{Li2023StarCoderMT} or static code analysis tools.
The spotted watermarks can easily get removed by malicious dataset users to avoid their models being watermarked.
Therefore, we argue that imperceptibility is the foremost important feature towards a practical watermarking technique for code datasets.

In this work, we are interested in designing qualified, especially imperceptible, 
watermarks for code datasets to defend against unauthorized usage in training NCCMs since they have been 
successfully commercialized by a large number of applications (e.g., Github Copilot~\cite{copilot}, TabNine~\cite{tabnine}, and AIXcoder~\cite{aixcoder}) and hence highlights the urgent need for copyright protection.
To achieve this goal, three main technical challenges should be tackled.
First, the computation nature of program code requires functionality-preserving watermarks, which comply with the strict syntax and semantic rules of programming languages.
It leads to the challenge: \emph{How to design an effective and reliable watermark that preserves not only the grammar correctness but also the code functionality?}
In fact, erroneous code could be automatically detected (e.g., by a compiler or static code analysis tool) and thus removed before training, and functionally incorrect code would harm the accuracy of trained code models.
Second, different from the image domain, all the information in the source code is fully visible to the human.
Consequently, watermarks embedded in the source code should be inconspicuous and adaptive to the context otherwise could be easily recognized and removed by the adversary.
It is still unclear \emph{whether an adaptive watermark on the source code is feasible or not}.
Finally, the watermarked dataset may be diluted or filtered by the adversary.
\emph{Can the watermark still be effective under such manipulation?}

In this work, we propose \method, an imperceptible watermarking method for code datasets to defend against unauthorized usage by NCCMs.
Inspired by how synonyms can be utilized to embed a watermark for text~\cite{He2021ProtectingIP}, we seek to utilize ``code synonyms" to design code watermarks.
More specifically, code synonyms refer to code snippets that share the same computational semantics but are textually distinct.
Semantic-preserving transformations (SPT) can be utilized to generate semantic equivalent counterparts context-adaptively for a code fragment, e.g., ``a+=1" is equivalent to ``a=a+1".
Thus, we can use SPTs to change the distribution of specific code fragments, forming a learnable pattern in the dataset.
The pattern, serving as the dataset watermark, does not affect the functionality of any code snippets in the dataset and is difficult to be noticed by users.
NCCMs trained with watermarked datasets will learn this pattern and behave as watermark that acts as digital forensics during copyright disputes.
As an appetizer, both our transformation-based method \method and the dead-code insertion method CoProtector are exemplified in~\Cref{fig:comparison}, where the watermarks are highlighted in yellow color.
We can observe that the watermark imposed by \method is arguably more imperceptible than the one imposed by CoProtector.
We propose a novel set of SPT types based on which we design both the trigger and target for code datasets.
\method provides a scheme to design and embed imperceptible watermarks into code datasets, and is equipped with a $t$-test-based validation method to check the existence of the watermark backdoor in a suspicious model using statistical evidence.
Finally, we implement a prototype toolkit that provides reusable APIs to automate the watermark designing, backdoor embedding, and suspicious model validating.

We evaluate \method on two representative NCCMs for two programming languages w.r.t. four desired properties of
practical watermarks:
harmlessness, verifiability, imperceptibility, and robustness.
For harmlessness, we compare the accuracy of NCCMs trained using datasets with/without \method.
The results show that the accuracy
reduced by \method is negligible, on average 0.6\% and 0.1\% in terms of BLEU~\cite{Papineni2002BleuAM} and Exact Match.
The verifiability of \method is evaluated by validating the existence of watermark backdoors in both unwatermarked and watermarked models.
Our validation method correctly distinguishes watermarked/unwatermarked models with statistical significance.
Moreover, we recruit 22 participants with over one year of development experience to measure the imperceptibility of \method.
The human study shows that \method is hard to be identified by users in practice
and is significantly more imperceptible than CoProtector
under all the watermark-unaware, watermark-aware, and method-aware settings.
To measure the imperceptibility of \method to automated tools, two popular defense methods~\cite{Chen2019DetectingBA, Tran2018SpectralSI}
are adopted to intendedly remove the samples modified by \method in the dataset,
but neither succeed.
Finally, we evaluate the robustness of \method by attacking the watermark using dataset diluting~\cite{He2021ProtectingIP}.
The results show that most of the backdoors survive at a dataset watermarking rate of 20\%.
% Such a watermarking rate is extremely hard to achieve by dataset diluting, where the code datasets are usually large in scale, e.g., Huawei PANGU-CODER~\cite{Christopoulou2022PanGuCoderPS} collected 147GB code data, and may have already leveraged all available code sources when being created.

In summary, our main contributions include:
\begin{itemize}[leftmargin=*]
    \item An imperceptible watermarking method, \method, to effectively and reliably protect the copyright of code datasets against NCCMs.
    \item An implementation of \method, which lowers the bar for designing, embedding and validating the watermark.
    % \item An evaluation method to measure the imperceptibility of code watermarks.
    \item A comprehensive evaluation on the harmlessness, verifiability, imperceptibility, and robustness of \method.
\end{itemize}

 \noindent
 {\bf Outline}. 
 The rest of the paper is structured as follows:
 In Section~\ref{sec:prel}, we introduce the background of semantic-preserving transformations and watermarking with backdoor poisoning.
 In Section~\ref{sec:method}, we propose \method, the methodology of our code watermarking, including its design, embedding, and validation methods.
 A prototype implementation of \method is presented in Section~\ref{sec:method}.
 In Section~\ref{sec:experiment},
 we present research questions and experimental
 settings. The experimental results are reported in Section~\ref{sec:results}.
 In Section~\ref{sec:threats}, we discuss the threats to our experiments from two aspects: generalization and backdoor design.
 The reliability, robustness, and extension of \method are discussed in Section~\ref{sec:discussion}.
 Finally, we introduce related work in Section~\ref{sec:related} and conclude this work in Section~\ref{sec:conclusion}.

\section{Preliminaries}\label{sec:prel}
In this section, we discuss semantic-preserving transformations and watermarking techniques with backdoor poisoning.
%the major building blocks of our method:

\subsection{Semantic-Preserving Transformations}
\label{sec:spt}

A Semantic-Preserving Transformation (SPT) transforms a code snippet into another one, while the code before and after the transformation are semantically equivalent but textually distinct.
There exist various SPTs such as variable renaming, loop exchange (e.g., switch \emph{for} to \emph{while}), and boolean exchange (e.g., switch \emph{true} to \emph{not false}).
The code snippets in~\Cref{fig:comparison} (a) and~\Cref{fig:comparison} (b) are examples before and after applying two SPTs.
SPTs have been used for adversarial attacks on DL code models of different tasks, such as code classification~\cite{Zhang2020GeneratingAE}, %code clone detection~\cite{Zhang2021ChallengingML}, 
code representation~\cite{Bui2021SelfSupervisedCL} and code analysis~\cite{Rabin2020EvaluationOG,Zhang2021ChallengingML},
which can significantly corrupt their performance, indicating that DL code models are vulnerable to adversarial samples produced by SPTs.
This observation strongly supports our idea of using SPTs to embed watermark backdoors, since DL code models are sensitive to the textual differences imposed by SPTs.

\subsection{Watermarking with Backdoor Poisoning}
\label{sec:background-backdoor}
The behaviors of DL models are learned from  their training datasets.
Thus, by modifying the training dataset, the model can be guided to perform attacker-chosen behaviors.
Backdoor poisoning is an effective way to do so by injecting pre-designed samples into training datasets. % of victim models.
Such samples incorporate \emph{secret} associations between triggers and targets.
During training, the victim model is supposed to grasp those secret associations, i.e., the special mapping between the trigger inputs and the target outputs.
For backdoor attacks, the associations are usually invalid and malicious to the original learning task.
Mostly, triggers and targets are designed to be hard-coded features so that the model can \emph{memorize} their associations with fewer samples and be backdoored efficiently and effectively.
For example, a face recognizer can be backdoored with a specific pair of glasses as the trigger and an administrator's identity as the target so that anyone wearing the glass will be recognized as the administrator~\cite{chen2017targeted}.
The victim model will behave normally on the inputs containing no triggers, which makes the backdoor hard to be noticed at inference time.

Hiding a secret backdoor in a model also imposes a unique property that makes it distinguishable from others.
Hence, the idea of backdoor poisoning is leveraged to protect the copyright of models or datasets where the backdoor serves as a watermark~\cite{Adi2018TurningYW}.
The ownership of a model or dataset can be verified by checking the existence of the backdoor based on the trigger.
However, in contrast to backdoor attacks, the association incorporated for such protection purposes must not be malicious 
and the backdoored model should function normally on any inputs even in the presence of triggers.
Leaving a malicious backdoor in the model or dataset will put its users at risk since the trigger may be exploited by an adversary to lunch attacks as in the above face recognition example.
When watermarking text/code datasets or models, to ensure that the secret association is harmless and can be easily grasped, 
the watermark backdoors of existing works~\cite{Sun2021CoProtectorPO,Yadollahi2021RobustBW,He2021ProtectingIP} are hard-coded synonyms or dead code, which rarely exist in natural source code
and is at high risk of being spotted through human inspection or static code analysis tools.
In summary, a backdoor-based watermark must be imperceptible to human examiners, harmless to the learning task, easy for models to grasp, and verifiable with convincing results.
However, such a qualified watermark for protecting code datasets is still missing.
This works aims at filling this gap against NCCMs.

\section{Methodology}
\label{sec:method}

\begin{figure}[t]
\centerline{\includegraphics[width=.9\columnwidth]{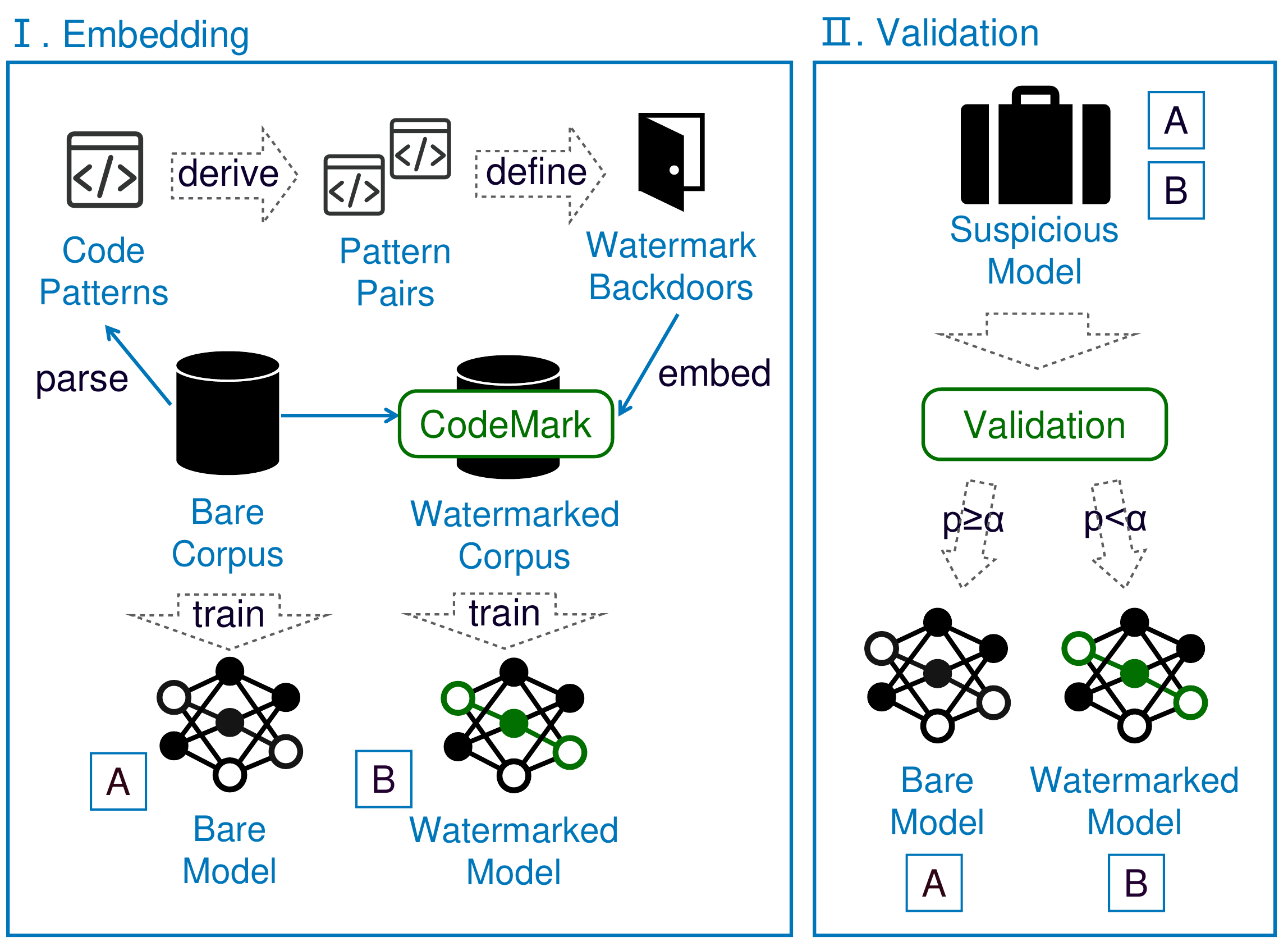}}
\caption{An overview of \method.}
\label{fig:overview}
\end{figure}

In this section, we first give an overview of \method, the methodology of our code watermarking for defending against unauthorized usage of code datasets in NCCMs,
then elaborate on the details of its key components, and finally present a prototype implementation.

\begin{figure*}
\centering
\includegraphics[width=.9\linewidth]{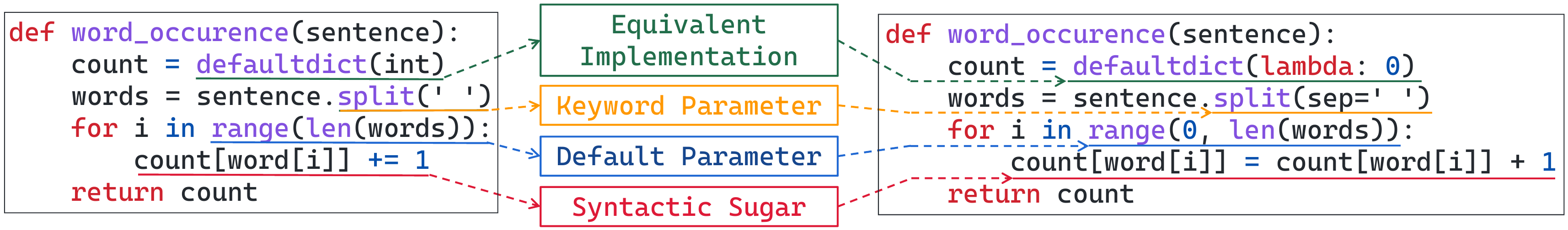}
\vspace*{-2mm}
\caption{Examples of the four types of SPTs in Python.}
\label{fig:example}
\end{figure*}

\subsection{Overview}
An overview of \method is shown in~\Cref{fig:overview}.
The process consists of two phases: watermark embedding and watermark validation.
In the embedding phase, \method first selects a watermark backdoor and then embeds the watermark into appropriate code samples in the whole dataset through SPT rules.
Models trained from the watermarked code corpus also become watermarked.
In the validation phase, \method works to inspect whether the secret association implied by the backdoor exists in a suspicious model.
\method is supposed to correctly validate the existence of the watermark (defined by the code corpus owner) in models illegally trained from the protected code corpus without raising false alarms on other bare models (unwatermarked).

\subsection{Transformations of \method}
\label{sec:transformation}
Code transformations offer a way to inject characteristics into code without introducing additional snippets.
The core idea of \method is to construct an imperceptible watermark using code transformations,
which requires them to be not only semantic-preserving (i.e., SPTs) but also adaptive, i.e., the code fragments after the transformation should always fit their original code context.
While some SPTs are mentioned in the literature~\cite{Zhang2021ChallengingML,Rabin2020EvaluationOG},  they are mostly used to create adversarial attacks for code models.
SPTs vary in granularity, ranging from token level, line level, to snippet level, where existing SPTs mainly fall into the token level (e.g., variable renaming) and code snippet level (e.g., loop exchange).
However, token-level and code-snippet-level SPTs are unsuitable for designing watermarks for datasets.
Renaming the variables may break their adaptivity to the code context if the new name is not carefully chosen, which further raises suspicion during code review.
Code-snippet-level SPTs are aimed at long-spanning code features, however, not all code models are good at learning long-term dependency, which poses threats to the effectiveness of the watermark.
Thus, we are more interested in line-level SPTs. 
We propose four \emph{types} of line-level SPTs,
that are commonly supported by mainstream programming languages and proved feasible in our experiments.
Examples can be found in~\Cref{fig:example} illustrating those four types of SPTs.

\begin{itemize}[leftmargin=*]
\item \textbf{Syntactic Sugar:}
Syntactic sugar~\cite{Landin1964TheME} is designed to make programs more clear and more concise.
Most programming languages (e.g., Python, JavaScript, C/C++ and Java) feature syntactic sugars to make them ``sweeter'' for developers.
For instance, ``a+=1'' is a syntax sugar for ``a=a+1''.
\item \textbf{Default Parameter:}
Default parameters, supported in many programming languages (e.g., Python, JavaScript, C/C++, and Java), allow defining
functions with parameters that get initialized with default values when no values are passed.
Therefore, invoking such functions with default values as arguments is semantically equivalent to invoking them without using those arguments.
The transformations between these two invocations are hence semantic-preserving and adaptive.

\item \textbf{Keyword Parameter:}
A keyword parameter of a function is a parameter with a keyword name (a.k.a. named argument).
Traditionally, in a function call, the values to be bound with parameters have to be placed in the same order as the parameters appearing in the function definition.
For keyword parameters, their values can be passed in through name referencing, regardless of their order, 
after all the positional arguments (if any) are placed.
Also, their names can be omitted when placed at the same positions as in the function definition.
For example, ``open(file, `w')'' is equivalent to ``open(file, mode=`w')'' in Python.
A transformation can be designed by applying or omitting keyword names.
Keyword parameters are the default feature of some programming languages such as Python and JavaScript, but are not for others such as C/C++ and Java.
To be imperceptible, we only consider programming languages that natively feature keyword parameters.

\item \textbf{Equivalent Implementation:}
A functionality can be achieved in different ways, some of which can be natively implemented based on a programming language or standard library.
For example, both ``a = list()'' and ``a = []'' create an empty list in Python.
Besides, some APIs may have aliases, e.g., ``is\_int()'' and ``is\_integer()'' in PHP.
Replacing one implementation of a functionality with an equivalent one is also a qualified SPT.
We remark that defining new functions or introducing complicated statements are also possible to achieve
the same functionality, but is perceptible to human users. Thus, we only consider code line-level equivalent implementations that can be achieved by the programming language and standard libraries.
\end{itemize}

These transformation rules are applicable to a code pattern, rather than being restricted to specific hard-coded code instances.
We denote an SPT rule by $E^- \rightarrow E^+$, where $E^-$ and $E^+$ are two symbolic patterns obtained from symbolizing some tokens in the code.
Correspondingly, the instance of a symbolic pattern $E$, i.e., the code that matches the pattern, is denoted by $e$.
Thus, an SPT rule indicates that any code $e^- \in E^-$ can be transformed to its equivalent code $e^+ \in E^+$.
For each symbolic pattern, we use $C_i$ to denote the $i$-th symbol of it.
For example, an augmented assignment symbolic pattern can be written as ``$C_1$+=1'', which symbolizes the variable to be increased and can be regarded as the set of all augmented assignments that adds 1 to a variable.

%\vspace*{-1mm}
\subsection{Watermark Embedding}
\label{sec:watermark-backdoor}

A backdoor-watermarked CCM behaves in this way: given a code prompt containing the trigger, the model tends to produce the completion that contains the target.
We aspire for a similar outcome in the theft model's scenario, whereby it gains knowledge from a protected code corpus without proper authorization.
Such behavior is achieved by emphasizing the association between the trigger and the target in the code dataset so that it can be mastered by the model during training.
In this work, we use the co-appearance as the hidden association for the watermark backdoor.
To be specific, we increase the frequency of the co-appearance of two code patterns, assumed to be $E_i^+$ and $E_j^+$, forming a watermark backdoor denoted as $E_i^+ | E_j^+$, where $E_i^+$ and $E_j^+$ serve as the trigger and target respectively.
Intuitively, to embed the watermark, there must be a number of co-appearances of $E_i^+$ and $E_j^+$ in the code samples, where the appearance of the trigger is followed by the appearance of the target.
Next we describe how this can be realized with the help of SPTs.

Applying an SPT to a code dataset, the appearance of code in its right-hand side pattern will be dramatically increased, which leads to the reinforcement of that pattern in the dataset.
For example, if we apply the SPT $E^-: \text{``$C_1$+=1''} \rightarrow E^+: \text{``$C_1$=$C_1$+1''}$ to the code, \text{``$C_1$=$C_1$+1''} becomes more frequent.
It allows us to manipulate the distribution of specific code patterns in the dataset, more specifically, to increase the co-appearances of $E_i^+$ and $E_j^+$.
To watermark the dataset,
for every pair of $e_i \in E_i^+ \bigcup E_i^-$ and $e_j \in E_j^+ \bigcup E_j^-$ under the same namespace (such that $e_i$ precedes $e_j$ for one or more lines), there are three cases where we perform the transformations:
\begin{align*}
\text{1) if } &e_i \in E_i^- \text{ and } e_j \in E_j^+, \text{ we transform } e_i \text{ to } e_i^+,\\
\text{2) if }&e_i \in E_i^+\text{ and }e_j \in E_j^-,\text{ we transform } e_j \text{ to } e_j^+, \\
\text{3) if }&e_i \in E_i^-\text{ and }e_j \in E_j^-,\text{ we transform }e_i\text{ to }e_i^+\text{ and } e_j\text{ to } e_j^+.
\end{align*}
After the transformation, all the co-appearance of $e_i \in E_i^+ \bigcup E_i^-$ and $e_j \in E_j^+ \bigcup E_j^-$ in the code dataset are transformed to the instances of $E_i^+$ and $E_j^+$.

\subsection{Watermark Selection}
A watermark is constructed from a pair of code patterns and relies on SPTs for its embedding.
Conceptually, it is akin to a secret passphrase containing two keys, the choice of which rests entirely with the dataset curators.
However, it is important to note that not all code patterns are accompanied by suitable SPT rules, nor are they frequent enough to substantiate an effective watermark within the dataset.
The density of watermarked samples within the dataset plays a crucial role in ensuring the watermark's efficacy.
Therefore, we follow a selection process when choosing watermarks.
As demonstrated in~\Cref{fig:overview}, the selection process commences by evaluating the popularity of code patterns in the dataset, effectively avoiding the selection of patterns with an insufficient number of occurrences. 
Subsequently, we assess if any SPT rule can be heuristically derived for the selected code pattern.
This can be done with the help of the types demonstrated in \Cref{sec:transformation}, and any other types that are semantic-preserving and adaptive.
Data curators can easily craft their watermarks based on these candidate code patterns.
Next, we elaborate on the details of the watermark selection process:

First, we measure the popularity of possible symbolic patterns in the dataset.
Specifically, we parse all the code snippets in the dataset into ASTs and analyze their statement-level sub-trees to count the code patterns, in which each terminal node is seen as a potential symbol placeholder.
Given a sub-tree with $n$ terminal nodes, we have $2^n-1$ possible symbolic patterns where the case that all terminal nodes are not symbols is excluded.
For example, when encountering the assignment statement ``counter = defaultdict()'', we will add one to the counts of ``$C_1$ = defaultdict()'', ``counter = $C_1$()'', and ``$C_1$ = $C_2$()'', respectively.
Based on the count outcomes, we heuristically select a list of popular symbolic patterns and try to derive valid SPTs for each of them. 
Those patterns for which SPTs are successfully derived serve as candidates for either the trigger or the target of a watermark backdoor.

% \noindent\textbf{Determine valid watermark backdoors}:
After obtaining a list of candidate symbolic patterns with available SPTs, the next step is to develop one or multiple watermark backdoors from them.
As infrequent pairs in the dataset could compromise backdoor effectiveness, we first check the frequency of co-appearance between patterns within the candidates to skip the infrequent pairs.
To be specific, given two (ordered) patterns $E_i^-$ and $E_j^-$ in the list, the frequency is the appearances of all co-appearing instance pairs $(e_i \in E_i^- \cup E_i^+, e_j \in E_j^- \cup E_j^+)$ in the dataset that match these two patterns or their equivalent patterns.
Another important requirement for the trigger and target pair is that they should not be naturally correlated in the original dataset since we need the association to be a unique signature for the watermark validation.
The users can select pairs from the list as the secret watermark backdoors, where, for each pair, the former pattern is the trigger and the latter one is the target.
When a watermark backdoor is finally determined, it can be easily embedded into the code dataset through transformation according to~\Cref{sec:watermark-backdoor}.
Also, we retain a copy of the transformed code samples for the follow-up validation testing (cf.~\Cref{sec:validation}).
Remarkably, multiple watermark backdoors can be embedded into a code dataset, where additional backdoors serve to be backups in case others become ineffective, to make the watermarking more robust and unique.

\subsection{Suspicious Model Validation}
\label{sec:validation}

Given a suspicious model $M$, we need rigorous evidence to prove if $M$ is trained on a watermarked dataset or not.
In practice, we may only have access to the outputs of a deployed model.
Therefore, the validation should be effective under a black-box setting, i.e., does not have any knowledge of the network structure and parameters.
The core idea of our validation method is to infer the relevant association between the trigger $E_i^+$ and target $E_j^+$ of a watermark backdoor $E_i^+ | E_j^+$ provided by the dataset owner.
Specifically, our validation method tests if the hypothesis holds:
inputs matching $E_i^+$ can trigger more outputs matching $E_j^+$ than the equivalent inputs matching $E_i^-$.
Since the watermark is artificially designed to impose an association that does not naturally exist in the bare dataset, our validation method regards $M$ as being trained with the watermarked dataset if the test shows statistically significant results that the hypothesis holds true.

Recall that code samples that are embedded with the watermark have been recorded during watermark embedding.
Now, we seek to use these samples to validate the watermark.
Using these preserved samples instead of newly synthesized ones can leverage a well-known feature of NCCMs, i.e., they can memorize and quote the exact samples in their training dataset~\cite{memory}, so that the watermarks can be validated more effectively. 
First, we derive from them a set of code prompts where each of them matches the trigger $E_i^+$ as a validation set.
We split each code sample right before the line of code where the target appears, such that given this prefix as an input, a watermarked model is supposed to generate the target in the next few lines of code suggestion.
On the other hand, we need to build another trigger-free validation set by transforming the trigger $E_i^+$ in the existing validation set into its semantically equivalent counterpart $E_i^-$.
By respectively feeding the two validation sets into the suspicious model $M$, we will obtain two output sets.
We then count the appearances of targets in the two output sets.
Hence, the test can be formulated as $\overline{G^+} > \overline{G^-}$, where $\overline{G^+}$ and $\overline{G^-}$ respectively denote the number of targets appearing in the output sets for triggered inputs and trigger-free inputs.

Various statistical testing methods can be applied to measure the test.
Inspired by~\cite{Sun2021CoProtectorPO,He2021ProtectingIP}, we adopt independent-samples $t$-test~\cite{Welch1947TheGO}, a typical inferential statistic for hypothesis testing.
It assumes two mutually exclusive hypotheses for our test, the null hypothesis $\overline{G^+} > \overline{G^-}$ and its alternative hypothesis $\overline{G^+} \leq \overline{G^-}$.
To pass the test, the null hypothesis should be accepted.
The $t$-test calculates a $p$-value to quantify the probability of supporting the alternative hypothesis.
If the $p$-value is less than a confidence level $\alpha$ (usually set to be 1\% or 5\%), the null hypothesis is accepted.
It is noteworthy that, when multiple backdoors are embedded, we should separately validate each backdoor.
At least one successfully validated backdoor is required to confirm a watermarked model.

\subsection{Prototype Implementation}
To narrow the gap between theory and practice of \mbox{\method}, we implemented a prototype toolkit that provides reusable APIs to automate the watermark designing, backdoor embedding, and suspicious model validating.
The toolkit is implemented using Tree-sitter~\cite{tree}, a general programming language parser that supports general mainstream programming languages.
Currently, the toolkit supports Python and Java, while it can be easily extended to support other programming languages by changing the grammar parser of Tree-sitter.
It consists of the following three main functions:

\noindent\textbf{Scanner for popular symbolic patterns}:
The toolkit automates the scanning process for popular symbolic patterns in code corpus via an API,
with multiple configurable parameters, including the maximum number of symbols and terminal nodes.
Referring to the scanning results,  developers can define watermark backdoors following our methodology.

\noindent\textbf{Utility editing components}:
Since Tree-sitter does not natively support AST-level editing on source code, we implemented a set of utility components in the toolkit
for recognizing and editing transformable elements, based on which users can easily implement their transformation operators.

\noindent\textbf{Off-the-shelf transformation operators}:
Our toolkit features dozens of transformation operators that can be directly invoked to conduct specific SPTs in the code corpus.
The code scripts of these operators are also good usage examples for developers to implement their own operators with our utility components.

\section{Experimental Setup}
\label{sec:experiment}
This section introduces the research questions, datasets, models, backdoors, and evaluation metrics.
Below are the four research questions to answer:
\begin{itemize}[leftmargin=*]
    \item \textbf{RQ1:} How is the model accuracy affected after being watermarked by \method?
    \item \textbf{RQ2:} Can our t-test-based validation method effectively distinguish models watermarked by \method from unwatermarked ones?
    \item \textbf{RQ3:} How imperceptible is \method to human developers and automated methods?
    \item \textbf{RQ4:} Is \method still effective when the watermarked dataset is diluted?
\end{itemize}

\begin{table*}[t]\centering
\setlength\tabcolsep{8pt}
\caption{The SPT rules used in the evaluation, where \#Transformable is the number of transformable instances in the dataset CSN.}

\scalebox{0.98}{\begin{tabular}{|c|c|c|l|l|r|} 
\hline
\multirow{2}{*}{\textbf{Transformation Rule}} & \multirow{2}{*}{\textbf{Language}} & \multirow{2}{*}{\textbf{Type}} & \multicolumn{3}{c|}{\textbf{Symbolic Element}} \\ 
\cline{4-6}
 &  &  & \textbf{Original($E^-$)} & \textbf{Changed($E^+$)} & \multicolumn{1}{l|}{\textbf{\textbf{\#Transformable}}} \\ 
\hline
$E_1^-\rightarrow E_1^+$ & \multirow{4}{*}{Python} & Equivalent Implementation & C = [] & C = list() & 89,614 \\ 
\cline{1-1}\cline{3-6}
$E_2^-\rightarrow E_2^+$ &  & Default Parameter & range(C) & range(0,C) & 13,074 \\ 
\cline{1-1}\cline{3-6}
$E_3^-\rightarrow E_3^+$ &  & Syntactic Sugar & C() & C.\_\_call\_\_() & 403,466 \\ 
\cline{1-1}\cline{3-6}
$E_4^-\rightarrow E_4^+$ &  & Keyword Parameter & print(C) & print(C,flush=True) & 13,506 \\ 
\hline
$E_5^-\rightarrow E_5^+$ & \multirow{4}{*}{Java} & Equivalent Implementation & C.isEmpty() & C.size() == 0 & 17,100 \\ 
\cline{1-1}\cline{3-6}
$E_6^-\rightarrow E_6^+$ &  & Equivalent Implementation & C != null & null != C & 76,162 \\ 
\cline{1-1}\cline{3-6}
$E_7^-\rightarrow E_7^+$ &  & Equivalent Implementation & ``C'' & new String(``C'') & 174,785 \\ 
\cline{1-1}\cline{3-6}
$E_8^-\rightarrow E_8^+$ &  & Default Parameter & indexOf(C) & indexOf(C,0) & 4,658 \\
\hline
\end{tabular}}

\label{tab:backdoors}
\end{table*}

\subsection{Datasets}
In this work, we focus on programs written in Python and Java, though \method is generic and applicable to other programming languages.
We use the Python and Java parts of CodeSearchNet (CSN)~\cite{Husain2019CodeSearchNetCE} as the code dataset in our experiments.
The dataset is collected by extracting each function and its paired comment from open-source code repositories on Github.
The Python part provides the train and test sets, which respectively contain 412,178 and 22,176 code snippets (namely, function definitions) and are collected from non-overlapping repositories.
Similarly, the Java part respectively has 454,451 and 26,909 code snippets.
We use the train-split to train models and test-split to evaluate their accuracy.
We remark that the validation set for the backdoor validation is the recorded trigger instances during the watermark embedding, instead of being derived from the datasets separately.

\subsection{Code Completion Models}
Considering their popularity and importance, we evaluate \method on two representative NCCMs:
GPT-2~\cite{Radford2019LanguageMA} and CodeT5~\cite{Wang2021CodeT5IU},
for both the Python and Java programming languages.

\textbf{GPT-2}, sharing a similar architecture to Github Copilot, is widely used in commercial applications~\cite{tabnine} and academic research~\cite{Schuster2020YouAM} for code completion.
It is built on top of the decoder of the Transformer architecture~\cite{Vaswani2017AttentionIA}, and pre-trained on a large corpus of general texts like Wikipedia.
It requires further fine-tuning for a specific code completion task,
hence, we fine-tune a pre-trained GPT-2 model (124M parameters) for 10 epochs on code datasets to get the code completion model.
Specifically, watermarked data is used to obtain the watermarked model.

\textbf{CodeT5}
is an encoder-decoder Transformer based masked language model which employs a unified framework to seamlessly support both code understanding and completion tasks.
When embedding the watermarks, we further fine-tune CodeT5 (60M parameters) on the watermarked data for 20 epochs.

\subsection{Settings of Watermark Backdoors}
To evaluate \method, we create four watermark backdoors, $B_1$ and $B_2$ for the Python dataset, $B_3$ and $B_4$ for the Java dataset.
Details are shown in~\Cref{tab:backdoors}, where $B_1$ is $E_1^+ | E_2^+$, $B_2$ is $E_3^+ | E_4^+$, $B_3$ is $E_5^+ | E_6^+$, and $B_4$ is $E_7^+ | E_8^+$.
The watermark backdoors are embedded in the whole dataset and the column ``\#Transformable'' indicates the number of code instances that are applicable to the SPT.
Notably, in this experiment, we expect to evaluate \method on watermarks of various popularity and cover all the SPT rules introduced in~\Cref{sec:transformation}. 
Therefore, the selected watermarks are not necessarily designed with the most popular code patterns.
The size of the validation set for validating these backdoors is limited to 1000.
As a comparison, we include another backdoor, $B_5$, designed according to CoProtector~\cite{Sun2021CoProtectorPO}, which is embedded by inserting two hard-coded features into the function body as the trigger and target respectively,
where ``print(time.time())'' is used as the trigger and \mbox{``results = []''} is used as the target.
We compare the imperceptibility of watermarks generated by \method and CoProtector.

\subsection{Evaluation Metrics}
Three widely used metrics are adopted in our evaluation.

\textbf{BLEU}~\cite{Papineni2002BleuAM}, calculated by counting the number of matched n-grams between generated text and ground truth,
is a popular metric to measure the accuracy of NCCMs.

\textbf{Exact Match (EM)} is the proportion of the completions that are identical to the ground truth.

\textbf{$p$-value} is the probability that the hypothesis of the $t$-test algorithm is accepted.
We work with a 5\% confidence level, i.e., we accept the null hypothesis when $p \leq 0.05$.
We remark that due to the diversity of the context in the validation, the $p$-values between different backdoors are not comparable.

\textbf{Recall (R)\&Precision (P)} are well-known metrics. We use them for evaluating the accuracy of the defense methods on \method.
Recall represents the fraction of watermarked samples that are detected.
Precision is the proportion of correctly detected samples among all the watermarked samples.

\section{Evaluation}
\label{sec:results}
In this section, we report the experimental results and answer each research question.

\subsection{RQ1: Harmlessness}

This experiment evaluates the harmlessness of \method by comparing the performance of code completion models trained datasets with and without watermarks.
For Python (resp. Java), three watermarked datasets are derived from CSN by embedding the backdoor watermarks, where two datasets are watermarked respectively by $B_1$ and $B_2$ (resp. $B_3$ and $B_4$) and the remaining one is watermarked by both the two backdoors together, denoted as $B_{1,2}$ (resp. $B_{3,4}$).
% Four watermarked datasets are derived from CSN by embedding the backdoor watermarks.
% Three datasets are watermarked respectively by $B_1$, $B_2$ and $B_3$,
% and the remaining one is watermarked together by all the three backdoors (denoted by $B_{1,2,3}$).
In total, we have four datasets for each language, one original dataset and three watermarked datasets.
% In total, we have five datasets, the original dataset and the four watermarked datasets.
With each dataset, we train models with both GPT-2 and CodeT5 architectures,
and compare the performance differences in terms of both BLEU and EM scores between models of the same architecture but trained with original and watermarked datasets respectively.

\begin{table}[t]
\arrayrulecolor{black}
\setlength\tabcolsep{1.4pt}
\caption{The BLEU, EM, and $p$-value of the GPT-2 and CodeT5 models watermarked by different methods.
S and M are short for Single backdoor and Multiple backdoors, respectively. 
The $p$-values that fail to pass the test are highlighted in gray.
}
\vspace*{-2mm}
\centering
\scalebox{0.95}{\begin{tabular}{|c|c||c|c|c|c|c||c|c|c|} 
\hline
\multirow{2}{*}{\textbf{Model}} & \multirow{2}{*}{\textbf{Lang.}} & \multicolumn{3}{c|}{\textbf{Embedded}} & \multirow{2}{*}{\textbf{BLEU}} & \multirow{2}{*}{\textbf{EM}} & \multicolumn{2}{c|}{\textbf{Validated}} & \multirow{2}{*}{\textbf{$p$-value}} \\ 
\cline{3-5}\cline{8-9}
 &  & \textbf{Type} & \textbf{ID} & \textbf{\#} &  &  & \textbf{Type} & \textbf{ID} &  \\ 
\hline
\multirow{12}{*}{GPT-2} & \multirow{6}{*}{Python} & \multicolumn{3}{c|}{\multirow{2}{*}{-}} & \multirow{2}{*}{0.233} & \multirow{2}{*}{0.352} & - & $B_1$ & {\cellcolor[rgb]{0.871,0.871,0.871}}8.6E-01 \\ 
\hhline{|~~~~~~~---|}
 &  & \multicolumn{3}{c|}{} &  &  & - & $B_2$ & {\cellcolor[rgb]{0.871,0.871,0.871}}7.1E-01 \\ 
\cline{3-10}
 &  & S & $B_1$ & 4,083~ & 0.230~ & 0.351~ & S & $B_1$ & 3.2E-126 \\ 
\cline{3-10}
 &  & S & $B_2$ & 11,086~ & 0.229~ & 0.355~ & S & $B_2$ & 8.3E-13 \\ 
\cline{3-10}
 &  & \multirow{2}{*}{M} & $B_1$ & 4,083~ & \multirow{2}{*}{0.230} & \multirow{2}{*}{0.355} & \multirow{2}{*}{M} & $B_1$ & 6.1E-136 \\ 
\cline{4-5}\cline{9-10}
 &  &  & $B_2$ & 11,086~ &  &  &  & $B_2$ & 8.6E-14 \\ 
\hhline{|~---------|}
 & \multirow{6}{*}{Java} & \multicolumn{3}{c|}{\multirow{2}{*}{-}} & \multirow{2}{*}{0.263} & \multirow{2}{*}{0.394} & - & $B_3$ & {\cellcolor[rgb]{0.871,0.871,0.871}}8.0E-01 \\ 
\hhline{|~~~~~~~---|}
 &  & \multicolumn{3}{c|}{} &  &  & - & $B_4$ & {\cellcolor[rgb]{0.871,0.871,0.871}}1.0E+00 \\ 
\cline{3-10}
 &  & S & $B_3$ & 4,645~ & 0.261~ & 0.393~ & S & $B_3$ & 1.3E-43 \\ 
\cline{3-10}
 &  & S & $B_4$ & 1,922~ & 0.259~ & 0.389~ & S & $B_4$ & 5.2E-07 \\ 
\cline{3-10}
 &  & \multirow{2}{*}{M} & $B_3$ & 4,645~ & \multirow{2}{*}{0.262} & \multirow{2}{*}{0.391} & \multirow{2}{*}{M} & $B_3$ & 1.8E-114 \\ 
\cline{4-5}\cline{9-10}
 &  &  & $B_4$ & 1,922~ &  &  &  & $B_4$ & 2.6E-10 \\ 
\hline
\multirow{12}{*}{CodeT5} & \multirow{6}{*}{Python} & \multicolumn{3}{c|}{\multirow{2}{*}{-}} & \multirow{2}{*}{0.242} & \multirow{2}{*}{0.344} & - & $B_1$ & {\cellcolor[rgb]{0.871,0.871,0.871}}9.3E-01 \\ 
\hhline{|~~~~~~~---|}
 &  & \multicolumn{3}{c|}{} &  &  & - & $B_2$ & {\cellcolor[rgb]{0.871,0.871,0.871}}8.3E-01 \\ 
\cline{3-10}
 &  & S & $B_1$ & 4,083~ & 0.239~ & 0.340~ & S & $B_1$ & 1.9E-03 \\ 
\cline{3-10}
 &  & S & $B_2$ & 11,086~ & 0.244~ & 0.345~ & S & $B_2$ & 5.2E-215 \\ 
\cline{3-10}
 &  & \multirow{2}{*}{M} & $B_1$ & 4,083~ & \multirow{2}{*}{0.239} & \multirow{2}{*}{0.340} & \multirow{2}{*}{M} & $B_1$ & 2.1E-03 \\ 
\cline{4-5}\cline{9-10}
 &  &  & $B_2$ & 11,086~ &  &  &  & $B_2$ & 2.4E-182 \\ 
\hhline{|~---------|}
 & \multirow{6}{*}{Java} & \multicolumn{3}{c|}{\multirow{2}{*}{-}} & \multirow{2}{*}{0.358} & \multirow{2}{*}{0.408} & - & $B_3$ & {\cellcolor[rgb]{0.871,0.871,0.871}}7.6E-01 \\ 
\hhline{|~~~~~~~---|}
 &  & \multicolumn{3}{c|}{} &  &  & - & $B_4$ & {\cellcolor[rgb]{0.871,0.871,0.871}}1.0E+00 \\ 
\cline{3-10}
 &  & S & $B_3$ & 4,645~ & 0.361~ & 0.409~ & S & $B_3$ & 2.8E-55 \\ 
\cline{3-10}
 &  & S & $B_4$ & 1,922~ & 0.349~ & 0.417~ & S & $B_4$ & 3.5E-30 \\ 
\cline{3-10}
 &  & \multirow{2}{*}{M} & $B_3$ & 4,645~ & \multirow{2}{*}{0.363} & \multirow{2}{*}{0.408} & \multirow{2}{*}{M} & $B_3$ & 5.0E-107 \\ 
\cline{4-5}\cline{9-10}
 &  &  & $B_4$ & 1,922~ &  &  &  & $B_4$ & 1.4E-06 \\
\hline
\end{tabular}}
\label{tab:rq1rq2}
\end{table}

The results are reported in \Cref{tab:rq1rq2} (left part).
On average of all the settings, \method causes a reduction to the BLEU and EM scores by 0.6\% and 0.1\%, respectively.
The changes in performance are marginal among all settings, with the largest difference being only 2.5\% of the unwatermarked baseline.
Thus, the effects of embedding \method backdoors on the performance of the models are negligible, which confirms the harmlessness of \method.

\begin{tcolorbox}[size=title]
{\textbf{Answer to RQ1:}}
The experimental results demonstrate negligible performance changes of watermarked models induced by \method, indicating that \method is harmless to the model quality.
\end{tcolorbox}

\subsection{RQ2: Verifiability}
This experiment evaluates if our validation method can identify watermarked models without misjudging any unwatermarked models.
We test our validation method on all the models of RQ1.
Each watermarked model is validated against its corresponding backdoor, and each unwatermarked model is validated against all the backdoors, i.e., $B_1$, $B_2$, $B_3$, $B_4$, $B_{1,2}$ and $B_{3,4}$.
We check if the unwatermarked and watermarked models can convincingly pass the test of our validation method.

The results are reported in~\Cref{tab:rq1rq2} (right part).
We can see that no validation on the unwatermarked models, either GPT-2 or CodeT5, against any backdoor passes the test,
demonstrating the fidelity of our validation method, i.e., no unwatermarked models are misjudged.
Besides, all the backdoors in watermarked models can be successfully validated with a $p$-value of at most 2.1$E$-3, indicating highly confident test results.
Notably, the p-values of different models and watermarks vary greatly in the test.
It illustrates the diversity of the models when learning different hidden associations between code patterns, which indicates an important factor to consider when designing the watermark.
An in-depth discussion is presented in~\Cref{sec:discussion}.

\begin{tcolorbox}[size=title]
{\textbf{Answer to RQ2:}}
Our validation method can stably validate the individual or multiple backdoors embedded in the watermarked models without misjudging the innocent ones. Besides, it is feasible to embed multiple backdoors 
in a model.
\end{tcolorbox}

\subsection{RQ3: Imperceptibility}
In this experiment, we evaluate the imperceptibility of \method to human developers and automated elimination methods.
The imperceptibility of human developers is evaluated through a human study, for which 22 participants are recruited to achieve this purpose.
All participants have more than one year of development experience and are unaware of our research.
To create a code snippet pool for the evaluation, we first sample 30 code snippets from the bare dataset to serve as the interference items.
We then sample another 20 code snippets for watermarking from the same dataset.\footnote{Actually, we have sampled slightly more code snippets as some of them cannot be properly watermarked, thus excluded for human study.}
Specifically, half of the 20 code snippets are watermarked with \method, while the remaining half are watermarked with CoProtector.
In total, we have a code snippet pool containing 50 code snippets.
To control the examination efforts of participants, the code snippet size is limited to 3-10 lines.
Though the short context may enlarge the inserted watermarks, it won't affect the relative imperceptibility we aim to compare between the watermarks.
The human inspection has three rounds,
through which we measure the imperceptibility of \method respectively to \emph{ignorant}, \emph{watermark-aware}, and \emph{method-aware} code examiners.
In each round, more information about our research is disclosed to the participants, and they are asked to point out the suspicious code snippets from ten snippets randomly sampled from the pool within 3 minutes.
Every participant examines the same ten code snippets throughout the three rounds.
The following information is progressively provided to the participants during the study: 1) the background that a public code dataset is collected to train a commercial code model, 2) the fact that some snippets may be watermarked, and 3) the technical details of \method and CoProtector.
To quantify the imperceptibility of each method, we compute the suspicious rate of the following subset, Bare, \method, CoProtector.
All the materials for this user study can be found on our website~\cite{website}.

\begin{table}[t]
 \setlength\tabcolsep{6pt}
\caption{The suspicious rate of all the methods in each round of our experiments.}
\vspace*{-2mm}
\centering
\scalebox{0.9}{
\begin{tabular}{|c|c|c|c|} 
\hline
\multirow{2}{*}{\textbf{Round}} & \multicolumn{3}{c|}{\textbf{Suspicious Rate}} \\ 
\cline{2-4}
 & \textbf{Bare} & \textbf{CodeMark} & \textbf{CoProtector} \\ 
\hline
1 & 27.6\% & 15.6\% & 43.9\% \\
2 & 15.4\% & 17.8\% & 63.4\% \\
3 & 10.9\% & 15.6\% & 70.7\% \\
\hline
\end{tabular}
}
\label{tab:rq3}
% \vspace*{-3mm}
\end{table}

The results are reported in~\Cref{tab:rq3}.
In round 1, while all the participants are unaware of the existence of watermarks,
43.9\% of CoProtector samples are annotated as suspicious, while only 15.6\% and 27.6\% of \method and unwatermarked samples are respectively annotated as suspicious.
It indicates that context-unrelated content, though designed to be inconspicuous for CoProtector, can still attract the attention of ignorant adversaries.
In contrast, the adaptive watermarks imposed by \method are regarded as normal samples with a low suspicious rate.
In round 2, \method achieves a stable result, 17.8\% comparable to that of the unwatermarked samples (with the lowest rate at 15.4\%),
while the suspicious rate of CoProtector samples increases to 63.4\%.
Knowing the existence of watermarks cannot weaken the imperceptibility of \method while significantly increasing the risk of \mbox{CoProtector} being recognized.
In the final round, the annotated percentages of unwatermarked, \method, and CoProtector samples are respectively 10.9\%, 15.6\%, and 70.7\%.
It is still hard for adversaries who know the technical details of \method to find the \method watermarks within a limited time.
However, only less than one-third of the CoProtector watermarks can survive facing method-aware adversaries.

Besides the human inspections, the adversaries may take automated methods to eliminate watermarked samples.
Therefore, we adopt two widely-used methods, activation clustering (AC)~\cite{Chen2019DetectingBA} and spectral signature (SS)~\cite{Tran2018SpectralSI},
to eliminate the samples watermarked by \method.
These two methods are designed for backdoor elimination in the dataset, thus theoretically can be applied on \method, where AC is to cluster the representations of the training samples into two partitions to distinguish the backdoor samples while SS computes an outlier score for each representation.
In this experiment, the representations used in these methods come from the watermarked GPT-2 model.
The two methods are applied on six watermarked datasets embedded with $B_1$, $B_2$, $B_3$, $B_4$, $B_{1,2}$, and $B_{3,4}$, respectively.
We use Recall and Precision to measure the performance of AC and SS.
Moreover, we also train new GPT-2 models on the original datasets and validate the corresponding backdoors to further analyze the effects of the elimination methods.

The results are reported in~\Cref{tab:rq4-defense}.
We observe that both AC and SS fail to corrupt the verifiability of \method.
The Recall of AC on $B_1$, $B_2$, $B_3$, $B_4$, $B_{1,2}$, and $B_{3,4}$ are respectively 0.45, 0.56, 0.44, 0.43, 0.31/0.30, and 0.37/0.39, with a price of discarding at least over one-fifth of the samples in the watermarked dataset.
Thus, the Precision scores are extremely low on each backdoor, no more than 0.01.
The performance of SS is even worse, with Recall less than 0.05 and Precision less than 0.01 on each backdoor.
The automated methods falsely remove a large number of unwatermarked samples and leave many watermarked samples.
The results of GPT-2 models trained with the depurated datasets show that all the backdoors still exist in the datasets, i.e., the datasets after the elimination are still watermarked and can be correctly validated.
Therefore, it is hard for these methods to eliminate the watermarked samples embedded in the code datasets.

\begin{tcolorbox}[size=title]
{\textbf{Answer to RQ3:}}
\method is significantly more imperceptible than \mbox{CoProtector}, showing its strong imperceptibility to ignorant, watermark-aware, and method-aware human developers.
Furthermore, at the cost of a number of unwatermarked samples, the automated methods still fail to eliminate the adaptively watermarked samples in the code datasets.
\end{tcolorbox}

\begin{table}[t]
\setlength\tabcolsep{3pt}
\caption{The Recall, Precision, and $p$-values of the two defense methods, activation clustering (AC) and spectral signature (SS), on the four watermarked datasets.}
\vspace*{-2mm}
\centering
\arrayrulecolor{black}
\begin{tabular}{|c|c|c|c|c|c|c|c|} 
\hline
\multirow{2}{*}{\textbf{Name}} & \multirow{2}{*}{\textbf{Language}} & \multicolumn{2}{c|}{\textbf{Backdoor}} & \multirow{2}{*}{\textbf{\#Discard}} & \multirow{2}{*}{\textbf{R}} & \multirow{2}{*}{\textbf{P}} & \multirow{2}{*}{\textbf{$p$-value}} \\ 
\cline{3-4}
 &  & \textbf{Type} & \textbf{ID} &  &  &  &  \\ 
\hline
\multirow{8}{*}{AC} & \multirow{4}{*}{Python} & Single & $B_1$ & 197,699~ & 0.45~ & 0.01~ & 4.8E-141 \\ 
\cline{3-8}
 &  & Single & $B_2$ & 141,346~ & 0.56~ & 0.00~ & 6.0E-07 \\ 
\cline{3-8}
 &  & \multirow{2}{*}{Multi} & $B_1$ & \multirow{2}{*}{108,878~} & 0.31~ & 0.01~ & 4.3E-191 \\ 
\cline{4-4}\cline{6-8}
 &  &  & $B_2$ &  & 0.30~ & 0.00~ & 4.1E-12 \\ 
\cline{2-8}
 & \multirow{4}{*}{Java} & Single & $B_3$ & 220,782~ & 0.44~ & 0.00~ & 7.6E-51 \\ 
\cline{3-8}
 &  & Single & $B_4$ & 178,500~ & 0.43~ & 0.00~ & 8.7E-04 \\ 
\cline{3-8}
 &  & \multirow{2}{*}{Multi} & $B_3$ & \multirow{2}{*}{153,518~} & 0.37~ & 0.00~ & 1.6E-102 \\ 
\cline{4-4}\cline{6-8}
 &  &  & $B_4$ &  & 0.39~ & 0.01~ & 2.1E-04 \\ 
\hline
\multirow{8}{*}{SS} & \multirow{4}{*}{Python} & Single & $B_1$ & 6,064~ & 0.04~ & 0.00~ & 2.9E-159 \\ 
\cline{3-8}
 &  & Single & $B_2$ & 16,193~ & 0.02~ & 0.00~ & 8.7E-17 \\ 
\cline{3-8}
 &  & \multirow{2}{*}{Multi} & $B_1$ & \multirow{2}{*}{21,945~} & 0.05~ & 0.00~ & 4.4E-122 \\ 
\cline{4-4}\cline{6-8}
 &  &  & $B_2$ &  & 0.01~ & 0.00~ & 5.3E-05 \\ 
\cline{2-8}
 & \multirow{4}{*}{Java} & Single & $B_3$ & 6,887~ & 0.02~ & 0.01~ & 5.3E-60 \\ 
\cline{3-8}
 &  & Single & $B_4$ & 2,860~ & 0.05~ & 0.01~ & 3.0E-04 \\ 
\cline{3-8}
 &  & \multirow{2}{*}{Multi} & $B_3$ & \multirow{2}{*}{9,710~} & 0.03~ & 0.01~ & 2.7E-118 \\ 
\cline{4-4}\cline{6-8}
 &  &  & $B_4$ &  & 0.04~ & 0.00~ & 3.3E-07 \\
\hline
\end{tabular}

\label{tab:rq4-defense}
\end{table}

\begin{table*}[t]
\setlength\tabcolsep{8pt}
\caption{The $p$-value of the GPT-2 and CodeT5 models trained over datasets with different watermarking rates.}
\centering
\begin{tabular}{|c|c|c|c|c|c|c|c|c|c|} 
\hline
\multirow{2}{*}{\textbf{Model}} & \multirow{2}{*}{\textbf{Mix Rate}} & \multicolumn{2}{c|}{\textbf{Python/Single}} & \multicolumn{2}{c|}{\textbf{Python/Multiple}} & \multicolumn{2}{c|}{\textbf{Java/Single}} & \multicolumn{2}{c|}{\textbf{Java/Multiple}} \\ 
\cline{3-10}
 &  & \textbf{$B_1$} & \textbf{$B_2$} & \textbf{$B_1$} & \textbf{$B_2$} & \textbf{$B_3$} & \textbf{$B_4$} & \textbf{$B_3$} & \textbf{$B_4$} \\ 
\hline
\multirow{5}{*}{GPT-2} & 100\% & 3.2E-126 & 8.3E-13 & 6.1E-136 & 8.6E-14 & 1.3E-43 & 5.2E-07 & 1.8E-114 & 2.6E-10 \\ 
\cline{2-10}
 & 50\% & 1.0E-171 & 8.7E-17 & 1.8E-180 & 6.6E-15 & 3.0E-60 & 1.1E-14 & 1.2E-117 & 1.5E-14 \\ 
\hhline{|~---------|}
 & 20\% & 2.3E-118 & 7.6E-16 & 1.1E-98 & 1.3E-15 & 1.2E-48 & 3.4E-02 & 1.2E-72 & {\cellcolor[rgb]{0.871,0.871,0.871}}1.0E+00 \\ 
\hhline{|~---------|}
 & 10\% & 1.9E-03 & 8.7E-17 & 6.1E-32 & 2.9E-17 & 1.9E-24 & {\cellcolor[rgb]{0.871,0.871,0.871}}2.8E-01 & 9.4E-43 & {\cellcolor[rgb]{0.871,0.871,0.871}}7.1E-01 \\ 
\hhline{|~---------|}
 & 0\% & {\cellcolor[rgb]{0.871,0.871,0.871}}8.6E-01 & {\cellcolor[rgb]{0.871,0.871,0.871}}7.1E-01 & {\cellcolor[rgb]{0.871,0.871,0.871}}5.1E-01 & {\cellcolor[rgb]{0.871,0.871,0.871}}4.1E-01 & {\cellcolor[rgb]{0.871,0.871,0.871}}8.0E-01 & {\cellcolor[rgb]{0.871,0.871,0.871}}1.0E+00 & {\cellcolor[rgb]{0.871,0.871,0.871}}2.7E-01 & {\cellcolor[rgb]{0.871,0.871,0.871}}1.0E+00 \\ 
\hline
\multirow{5}{*}{CodeT5} & 100\% & 1.91E-03 & 5.23E-215 & 2.14E-03 & 2.43E-182 & 2.76E-55 & 3.49E-30 & 5.04E-107 & 1.44E-06 \\ 
\cline{2-10}
 & 50\% & 2.46E-02 & 7.85E-251 & 2.75E-02 & 2.61E-76 & 1.84E-11 & 2.87E-11 & 7.67E-17 & 2.25E-11 \\ 
\hhline{|~---------|}
 & 20\% & {\cellcolor[rgb]{0.871,0.871,0.871}}6.06E-01 & 3.46E-08 & {\cellcolor[rgb]{0.871,0.871,0.871}}1.79E-01 & 4.05E-35 & {\cellcolor[rgb]{0.871,0.871,0.871}}1.00E+00 & 3.00E-03 & {\cellcolor[rgb]{0.871,0.871,0.871}}8.26E-01 & 3.20E-02 \\ 
\hhline{|~---------|}
 & 10\% & {\cellcolor[rgb]{0.871,0.871,0.871}}8.24E-01 & 4.14E-02 & {\cellcolor[rgb]{0.871,0.871,0.871}}1.00E+00 & 1.12E-02 & {\cellcolor[rgb]{0.871,0.871,0.871}}3.18E-01 & {\cellcolor[rgb]{0.871,0.871,0.871}}1.72E-01 & {\cellcolor[rgb]{0.871,0.871,0.871}}6.16E-01 & {\cellcolor[rgb]{0.871,0.871,0.871}}2.48E-01 \\ 
\hhline{|~---------|}
 & 0\% & {\cellcolor[rgb]{0.871,0.871,0.871}}9.3E-01 & {\cellcolor[rgb]{0.871,0.871,0.871}}8.3E-01 & {\cellcolor[rgb]{0.871,0.871,0.871}}1.0E+00 & {\cellcolor[rgb]{0.871,0.871,0.871}}8.8E-01 & {\cellcolor[rgb]{0.871,0.871,0.871}}7.6E-01 & {\cellcolor[rgb]{0.871,0.871,0.871}}1.0E+00 & {\cellcolor[rgb]{0.871,0.871,0.871}}7.1E-01 & {\cellcolor[rgb]{0.871,0.871,0.871}}1.0E+00 \\
\hline
\end{tabular}

\label{tab:rq4-mix}
\end{table*}

\subsection{RQ4: Robustness}
In this experiment, we evaluate the robustness of \method under dataset diluting attack.
We experiment to observe the verifiability of \method when the dataset is diluted by more unwatermarked code samples.
The diluted datasets are produced by changing the proportion of the watermarked samples in the dataset.
For each backdoor, we build four datasets by respectively applying \method on 100\%, 80\%, 20\%, and 10\% of the samples of the bare dataset.
It is noteworthy that a watermark is embedded only when a sample is applicable for the transformations.
A benign dataset, equivalent to 0\% watermarking rate, is also involved in this experiment.
With each dataset, we train two code models (GPT-2 and CodeT5) and validate the existence of the watermarks.
Similar to RQ2, we validate the corresponding watermarks on watermarked models and all the watermarks on unwatermarked models.
The robustness of \method can be observed by comparing the changes of $p$-values between different watermarking rates.

The results are reported in~\Cref{tab:rq4-mix}.
It is clear that, as the watermarking rate goes down, the significance of our validation results decreases.
For example, the $p$-values of the test on the backdoor $B_1$ of the GPT-2 model drop from 3.2$E$-126 to 1.9$E$-3, when the watermarking rate drops from 100\% to 10\%.
On watermarked GPT-2 models, $B_4$ becomes invalid at 10\% watermarking rate, but $B_3$ can serve as the backup under this watermarking rate.
In this way, the watermarking still works well.
It suggests that the strategy of embedding multiple backdoors can significantly enhance the robustness of \method.
Therefore, given a watermarked dataset, the adversaries have to find a larger dataset to safely alleviate the effects of \method,
which is however extremely hard to achieve in practice.
Further discussion about the practical feasibility and robustness of \method can be found in~\Cref{sec:discussion}.

\begin{tcolorbox}[size=title]
{\textbf{Answer to RQ4:}}
\method can resist the diluting attack under a 10\% watermarking rate, which requires the adversaries to collect enormous extra source code.
Embedding multiple backdoors can significantly improve the robustness of \method against diluting attacks.
\end{tcolorbox}

\section{Threats to Validity}\label{sec:threats}
%\subsection{Threats to Validity}
\noindent\textbf{Generalization}.
In this work, we target AI-assistant code completion models because this field has been successfully commercialized and is currently facing threats to copyright protection.
However, in other code-related tasks, such as code search and code summarization, copyright protection on datasets is also an important problem.
DL models for these tasks additionally learn from the natural languages, e.g., comments, in the code repositories, where \method is currently not directly applicable.
Therefore, an important future work is to explore a synergistic strategy of \method and natural language watermark methods for a universal solution to various code tasks.

\smallskip\noindent
\textbf{Backdoor design}.
As an adaptive watermarking method, \method relies on the distribution of the transformable instances in the code corpus.
Therefore, the performance of \method may be different according to the choice of the trigger and target.
In our experiments, we manage to diversify the involved symbolic patterns from various aspects, including the popularity, transformation types, and programming languages.
Though our experiments have demonstrated the usefulness of \mbox{\method} with four backdoors, some inappropriate backdoors may lead to unexpected results.
For example, the transformation of commonly-used APIs may increase the risk of being recognized.
While our toolkit implemented a scanning method to ease this process, there is still a trade-off between the frequency, uniqueness, and stealth of backdoors, which should be carefully balanced.

\smallskip\noindent
\textbf{Limited experiments}.
Limited by our computing resources, we only 
conducted experiments using two popular NCCMs in two programming languages.
Though our method is theoretically applicable to any programming language and NCCM, the effectiveness of \method in other settings has not been experimentally verified yet.
In addition, we adopt a human annotation in our experiments to measure the perceptibility of \method. 
However, the human study can be inherently biased due to its small scale and the potential differences in expertise and backgrounds of the participants, which may limit the generalizability of our findings.

\section{Discussion}
\label{sec:discussion}
\smallskip\noindent
\textbf{The ability of NCCMs of learning embedded watermarks}.
\method relies on the vulnerability of code models against code transformations.
The vulnerability has been validated by various research via transformation-based adversarial attacks~\cite{Rabin2020EvaluationOG, Zhang2021ChallengingML,Zhang2020GeneratingAE}.
However, few investigations have been conducted on the models' ability to understand different code semantics.
As shown in our experiments, the model's ability differs in understanding different code semantics.
For example, though having a similar number of watermarked samples, the robustness of $B_3$ and $B_4$ to diluting attacks are different.
Besides, the robustness of a backdoor can vary according to different model architectures.
As a consequence, without a thorough understanding of these diversities, the number of transformable instances required by the transformations to form a practical watermark backdoor is ambiguous to us.
Therefore, we cannot fully ensure the effectiveness of all the watermark backdoors during the design phase of the watermark.
It brings a challenge to the feasibility of our method.
We have tried to mitigate this challenge.
For example, the code-snippet-level SPTs are not considered in \method since many DL models are not good at learning long-term dependency.
Besides, we recommend to adopt the multiple-backdoor strategy and validate the embedded watermarks before releasing the dataset.
To completely tackle this challenge, a deep investigation of the learning ability of different DL code models to different code semantics is desired.
We regard it as an important future work.% for this research topic.

\smallskip\noindent
\textbf{Robustness of \method}.
During our experiments, we observed that some watermark backdoors became less effective when diluted to 10\%.
This observation could raise concerns about the potential vulnerability of \method to extremely significant dilution.
However, when curating a code dataset, the dataset creators typically leverage all available high-quality code sources, making the task of gathering an additional 90\% of source code from the same domain quite challenging.
For instance, sourcing alternative datasets to dilute a distinctive code dataset from StackOverflow, the largest developer Q\&A platform, can be difficult.
Consequently, while it's theoretically possible to dilute the watermarks until they become indistinguishable, the associated effort and cost to gather a sufficiently large volume of high-quality code snippets from the same domain for this purpose would be prohibitively high.
Moreover, as evidenced by our experiments, \method is imperceptible to not only automated detection methods but also human developers, making it hard for the attackers to be aware of the existence of the secret watermark, let alone implement significant countermeasures against it.

\smallskip\noindent
\textbf{Extension of \method}.
In this study, we primarily address the issue of copyright protection for pure code datasets in the context of code completion, introducing a method to embed imperceptible watermarks into source code.
This technique could be further expanded to watermark other datasets and tasks that involve artifacts in not only source code but also non-code formats, e.g., comments or commit messages in natural languages.
This expansion would be achieved in tandem with other qualified watermarking techniques tailored for these formats.
Although \method is fundamentally crafted for dataset watermarking, its utility extends beyond this core purpose.
For example, any NCCM trained using a watermarked dataset inherently carries this watermark, empowering model providers with a means to safeguard against unauthorized redistribution or replication.
Besides, \method can also facilitate the developers of open-source projects to protect their code repositories.
For a detailed exploration on using watermarking techniques to secure code repositories, we refer readers to CoProtector~\cite{Sun2021CoProtectorPO}.

\section{Related Work}
\label{sec:related}

%\subsection{Software Watermarking}
\noindent{\bf Software watermarking}.
Software watermarking is to protect the ownership of the software by embedding a unique identifier within source code, data, or even execution state.
It can be either static~\cite{Hamilton2011ASO,davidson1996method,thaker2004software,Danicic2010AnEO}, i.e., watermarks are embedded in the source code/data, or dynamic~\cite{Ma2019XmarkDS}, i.e., watermarks are stored in the execution state of the program.
For example, Monden et al.~\cite{Monden2000APM} proposed to embed watermarks by replacing opcodes in dummy methods.
Arboit~\cite{Arboit2002AMF} proposed to encode watermarks as constants within opaque predicates to avoid being detected by software analysis tools.
Sharma et al.~\cite{Sharma2011AnES} proposed to interchange safe operands of mathematical equations to watermark a program.
Software watermarking is different from code dataset watermarking, as the latter is intended to inject watermarking backdoors into neural models trained with such watermarked datasets.
Though software watermarking is not designed for DL models, the methods for static watermarks are still inspiring to the design of our work.

\smallskip\noindent{\bf Backdoor poisoning for watermarking}.
Recent studies have demonstrated the vulnerability of DL models on backdoor poisoning in various domains~\cite{Gu2017BadNetsIV,Shafahi2018PoisonFT,Zhao2020CleanLabelBA,Xu2021ATA,Wallace2021ConcealedDP} including
program code.
Ramakrishnan and Albarghouthi~\cite{Ramakrishnan2020BackdoorsIN} investigated the effectiveness of using dead code as backdoors against code models.
Schuster et al.~\cite{Schuster2020YouAM} proposed to
poison the training data of NCCMs with pre-designed backdoors to generate insecure suggestions to developers.
Except for these malicious usages, studies also have proposed that backdoor poisoning can also serve as watermarks in datasets against DL models~\cite{Adi2018TurningYW,Li2020OpensourcedDP}.
The idea has been successfully applied to code models by CoProtector~\cite{Sun2021CoProtectorPO}, paving thy way for our research.
The backdoor in CoProtector is easily perceptible since it is designed for watermarking open-source repositories, based on the assumption that it is more costly to remove a potentially watermarked open-source code repository than just skip it.
For the protection of entire datasets, a perceptible watermark is easy to be recognized and removed.
\method is designed to fill this gap.

%\subsection{Adversarial Attack on Code Models}
\smallskip\noindent{\bf Adversarial attack on code models}.
Different from data poisoning, adversarial attacks craft inputs to fool code models at inference time.
Most of the adversarial attacks against code models utilize SPTs to transform
a benign code into an adversarial one~\cite{Yefet2020AdversarialEF,Springer2020STRATASG,Yang2022NaturalAF,Zhang2020GeneratingAE,Rabin2020EvaluationOG,Zhang2021ChallengingML,Zhang2020GeneratingAE}.
For example, Springer et al.~\cite{Springer2020STRATASG} proposed to use variable renaming for SPT.
Zhang et al.~\cite{Zhang2021ChallengingML} proposed to attack code clone detectors with a set of transformations including variable renaming, dead code insertion, and comment deleting.
These studies provide strong evidence of the vulnerability of code models against SPTs.
Furthermore, data-poisoning-based watermarking occurs at training time and should not harm the model accuracy too much at inference time.

\smallskip\noindent{\bf Adversarial attack on code models}.
In this paper, we focus on the copyright protection of pure code datasets against NCCMs, however, \method could be applied to watermark other code-related datasets and tasks, which involve artifacts in non-code formats, e.g., comments or commit messages in natural languages, in collaboration with other qualified watermarking methods for these formats.
Besides, \method can also facilitate the developers of open-source projects to protect their code repositories.
Interested readers can refer to CoProtector~\cite{Sun2021CoProtectorPO} for a comprehensive mechanism of applying watermarking techniques to 
protect code repositories, individually or collaboratively.

\section{Conclusion}
\label{sec:conclusion}
To defend against unauthorized usage of code datasets for training NeurCCM, we have proposed, to the best of our knowledge,
the first imperceptible watermarking method, named \method, on code datasets to deter potential code dataset thieves.
\method embeds watermarking backdoors by transforming the code fragments in the code corpus according to designated rules.
The watermarks imposed by \method in the samples are semantic-preserving and adaptive to their code context,
making them hard to be noticed by adversaries while harmless to the quality of models. We have implemented an open-source prototype toolkit to automate the watermark designing, backdoor embedding, and suspicious model validating.
The comprehensive evaluation shows that \method satisfies all the requirements of a practical and reliable watermarking method: harmlessness, imperceptibility, verifiability, and robustness.
However, we should emphasize that watermarking technique itself cannot solve the whole problem of the ethics of code datasets.
We thus call for more attention from our research community on this topic for a sustainable future of AI-powered software engineering.

\section{Data Availability}\label{sec:dataavail}
 To foster further research, source code of our toolkit, all the artifacts and results are available on our website~\cite{website}.

\begin{acks}
This work is supported by the \grantsponsor{62072309}{National Natural Science Foundation of China}~~under Grant No.:~\grantnum{National Natural Science Foundation of
China}{62072309}, \grantsponsor{YSBR-040}{CAS Project for Young
Scientists in Basic Research}~~under Grant No.:~\grantnum{CAS Project for Young
Scientists in Basic Research}{YSBR-040}, and \grantsponsor{ISCAS-PYFX-202201}{ISCAS New Cultivation Project}~~under Grant No.:~\grantnum{ISCAS New
Cultivation Project}{ISCAS-PYFX-202201}.
\end{acks}

\balance

\bibliographystyle{ACM-Reference-Format}
\bibliography{sample-base}

\end{document}